\documentclass[12pt]{article}
\usepackage{amsmath,amssymb,empheq,url} 
\usepackage[dvipsnames]{xcolor}
\usepackage[text={17cm,23cm}]{geometry}   

\def\red#1{{\color{red} #1}}

\let\OLDthebibliography\thebibliography
\renewcommand\thebibliography[1]{
  \OLDthebibliography{#1}
  \setlength{\parskip}{3pt}
  \setlength{\itemsep}{3pt plus 0.3ex}
}

\begin{document}
\def\prg#1{\medskip\noindent{\bf #1}}   \def\ra{\rightarrow}
\def\lra{\leftrightarrow}               \def\Ra{\Rightarrow}
\def\nin{\noindent}                     \def\pd{\partial}
\def\dis{\displaystyle}
\def\grl{{GR$_\Lambda$}}                \def\Lra{{\Leftrightarrow}}
\def\Leff{\hbox{$\mit\L_{\hspace{.6pt}\rm eff}\,$}}
\def\bull{\raise.25ex\hbox{\vrule height.8ex width.8ex}}
\def\ric{{Ric}}                         \def\tric{{(\widetilde{Ric})}}
\def\tmgl{\hbox{TMG$_\Lambda$}}         \def\tgr{{GR$_\parallel$}}
\def\Lie{{\cal L}\hspace{-.7em}\raise.25ex\hbox{--}\hspace{.2em}}
\def\sS{\hspace{2pt}S\hspace{-0.83em}\diagup}   \def\hd{{^\star}}
\def\dis{\displaystyle}                 \def\ul#1{\underline{#1}}
\def\mb#1{\hbox{{\boldmath $#1$}}}      \def\ir#1{\hspace{2.5pt}^{#1}\hspace{-1.5pt}}
\def\T{\mathbb{T}}         \def\tgl{{TG$_\lambda$}} \def\irr#1{\hspace{1pt}^{(#1)}\hspace{-1pt}}

\def\inn{\hbox{\vrule height0pt width4pt depth0.3pt
\vrule height7pt width0.3pt depth0.3pt
\vrule height0pt width2pt depth0pt}\hspace{0.8pt}}

\def\semidirect{\;{\rlap{$\supset$}\times}\;}
\def\bm#1{\hbox{{\boldmath $#1$}}}
\def\nb#1{\marginpar{{\bf #1}}\,}
\def\orth{{\perp}}
\def\tb{\hbox{\tiny$\bullet$\hspace{0.5pt}}}

\def\G{\Gamma}        \def\S{\Sigma}        \def\L{{\mit\Lambda}}
\def\D{\Delta}        \def\Th{\Theta}       \def\cP{{\cal P}}
\def\a{\alpha}        \def\b{\beta}         \def\g{\gamma}
\def\d{\delta}        \def\m{\mu}           \def\n{\nu}
\def\th{\theta}       \def\vth{\vartheta}   \def\l{\lambda}
\def\vphi{\varphi}    \def\ve{\varepsilon}  \def\p{\pi}
\def\r{\rho}          \def\Om{\Omega}       \def\om{\omega}
\def\s{\sigma}        \def\eps{\epsilon}    \def\t{\tau}
\def\nab{\nabla}      \def\bB{{\bar C}}     \def\bR{{\bar R}}
\def\heps{\hat\eps}   \def\tD{{\tilde\nab}} \def\bu{{\bar u}}

\def\bh{{\bar h}}     \def\bom{{\bar\om}}   \def\bg{{\bar g}}
\def\tphi{{\tilde\vphi}}  \def\tt{{\tilde t}} \def\hu{{\hat u}}
\def\tcV{{\tilde{\cal V}}} \def\hpi{{\hat\pi}}
\def\bm{{\bar m}}     \def\bn{{\bar n}}     \def\bb{{\bar b}}
\def\bi{{\bar\imath}} \def\bj{{\bar\jmath}} \def\bk{{\bar k}}

\def\tG{{\tilde\Gamma}} \def\cF{{\cal F}}    \def\bH{{\bar H}}
\def\cL{{\cal L}}     \def\bcL{{\bar\cL}}    \def\hL{{\hat L}}
\def\tcL{{\tilde\cL}} \def\cM{{\cal M }}     \def\cE{{\cal E}}
\def\cV{{\cal V}}     \def\hcV{{\hat\cV}}    \def\cH{{\cal H}}
\def\bcH{{\bar\cH}}   \def\hcH{\hat{\cH}}    \def\hH{\hat{H}}
\def\cA{{\cal A}}     \def\cT{{\cal T}}      \def\cB{{\cal B}}
\def\cK{{\cal K}}     \def\hcK{\hat{\cK}}    \def\cE{{\cal E}}
\def\cR{{\cal R}}     \def\hcR{{\hat\cR}}    \def\hR{{\hat R}{}}
\def\cO{{\cal O}}     \def\hcO{\hat{\cal O}} \def\tom{{\tilde\om}}
\def\tA{{\tilde A}}   \def\tT{{\tilde T}}   \def\tR{{\tilde R}}
\def\bT{{\bar T}}      \def\bP{{\bar P}}     \def\tL{{\tilde\L}}
\def\bG{{\bar G}}     \def\bS{{\bar S}}      \def\bH{{\bar H}}
\def\bK{{\bar K}}     \def\bL{{\bar L}}      \def\hcT{{\hat\cT}}
\def\hcR{{\hat\cR}}   \let\Pi\varPi          \def\bk{{\bar k}}
\def\bbf{{\bar f}}    \def\ct{\check\tau}    \def\vr{{\red{v}}}
\def\chH{{\check\cH}}           \def\mT{\hbox{$\mathbb{T}$}}
\def\fT{f(\mT)}                 \def\ub#1{\underbrace{#1}}

\def\rdc#1{\hfill\hbox{{\small\texttt{#1}}}}
\def\chm{\checkmark}  \def\chmr{\red{\chm}}  \def\vth{{\vartheta}}
\def\ccl{\hbox{$\vth$-$\om$-$\l$}}   \def\cc{\hbox{$\vth$-$\om$}}
\def\cLor{\hbox{$\vth$-$\L$}}

\def\nn{\nonumber}                    \def\vsm{\vspace{-9pt}}
\def\be{\begin{equation}}             \def\ee{\end{equation}}
\def\bega{\begin{align}}              \def\enda{\end{align}}
\def\bea{\begin{eqnarray} }           \def\eea{\end{eqnarray} }
\def\beann{\begin{eqnarray*} }        \def\eeann{\end{eqnarray*} }
\def\beal{\begin{eqalign}}            \def\eeal{\end{eqalign}}
\def\lab#1{\label{eq:#1}}             \def\eq#1{(\ref{eq:#1})}
\def\bsubeq{\begin{subequations}}     \def\esubeq{\end{subequations}}

\def\bitem{\begin{itemize}\vspace{-3pt} \setlength\itemsep{-3pt} }
  \def\eitem{\end{itemize}\vspace{-3pt} }
\renewcommand{\theequation}{\thesection.\arabic{equation}}
\def\ns#1{{\normalsize #1}}  

\def\rd{{\rm d}}
\def\tm{{\tilde m}}         \def\tn{{\tilde n}}         \def\tk{{\tilde k}}

\title{From the Lorentz invariant to the coframe form\\ of $\fT$ gravity}

\author{Milutin Blagojevi\'c\footnote{\texttt{mb@ipb.ac.rs}}\\
\ns{Institute of Physics, University of Belgrade,
                      Pregrevica 118, 11080 Belgrade, Serbia}\\[5pt]
James M. Nester\footnote{\texttt{nester@phy.ncu.edu.tw}}\\
\ns{Department of Physics, National Central University,
    Chungli 32001, Taiwan,}  \\[-3pt]
\ns{Graduate Institute of Astronomy, National Central University,
    Chungli 32001, Taiwan,}  \\[-3pt]
\ns{Leung Center for Cosmology and Particle Astrophysics}, \\[-3pt]
\ns{National Taiwan University, Taipei 10617, Taiwan} \\[-3pt]
\ns{and Center for Mathematics and Theoretical Physics,}   \\[-3pt]
    \ns{National Central University, Chungli 32001, Taiwan} }

\date{}
\maketitle

\begin{abstract}
It is shown that the Lorentz invariant $f(T)$ gravity, defined by the coframe-connection-multiplier form of the Lagrangian, can be gauge-fixed to the pure coframe form. After clarifying basic aspects of the problem in the Lagrangian formalism, a more detailed analysis of this gauge equivalence is given relying on the Dirac Hamiltonian approach.
\end{abstract}

\section{Introduction}
\setcounter{equation}{0}

Teleparallel geometry was introduced into physics by Einstein in the 1920s, in
an attempt to unify general relativity (GR) with electromagnetism \cite{sauer}. Although this goal has never been accomplished, the concept of teleparallel geometry was later revived by M\o ller \cite{moller} as a framework for defining gravitational energy, then it continued to live on as an arena for the pure gravitational dynamics. Nowadays, teleparallel gravity (TG) can be most naturally described as the gauge theory of translations, a subcase of the Poincar\'e gauge theory with vanishing curvature but nontrivial torsion \cite{hs-1979,hehl-1980}. In the absence of matter, there is a particular choice of the quadratic TG Lagrangian $\cL_T$, denoted as $\mT$, for which this theory becomes dynamically equivalent to GR; it is known as the teleparallel equivalent of GR \cite{mb-2002,mb.hehl-2012} (\tgr\ or TEGR).

Observational predictions of GR, as well as of its teleparallel equivalent \tgr, are highly successful not only at low energies  (solar system), but also in some high energy regimes \cite{GRtests}. On the other hand, there are some aspects of the cosmological dynamics, such as dark energy and dark matter, where convincing explanations within GR are missing. In such a situation, it has been quite natural to search for and investigate alternative gravitational theories \cite{modgr}. In the present paper, our attention is focused on $\fT$ gravity \cite{fT}, an extension of \tgr\ which is presently less understood than its GR counterpart, $f(R)$ gravity \cite{fR}.

As we know from GR, exact solutions of a gravitational theory are of essential importance for understanding its physical content. In particular, these solutions are naturally related to certain symmetry aspects of the theory. In the case of  pure coframe $\fT$ gravity, where the only dynamical variable is the {1-form} $\vth^i$, certain difficulties have existed in constructing exact solutions, caused by inappropriate ansatzes for the {coframes}\footnote{Lack of understanding the process of construction can be best illustrated by noting that dynamically incompatible ansatzes were often referred to as ``bad tetrads", in contrast to the ``good" ones.}. These difficulties were overcome by introducing the procedure of covariantization, in which the (Lorentz) covariant $\fT$ gravity is \emph{reconstructed} from its coframe form, see for instance Refs. \cite{krssak-2015,bejarano-2019,golovnev-2017,krssak-2018}. Nowadays, when the Poincar\'e gauge theory (PG) is known to be a well-established gravitational theory \cite{hehl-1980,mb-2002,mb.hehl-2012}, there is a more natural way to understand the result of the reconstruction procedure, based on treating any general TG as a subcase of PG, characterized by a vanishing curvature 2-form $R^{ij}=0$. As a consequence:
\bitem
\item the resulting general TG, defined in terms of the coframe $\vth^i$ and the flat (pure gauge) connection $\om^{ij}$, inherits the Lorentz and translational gauge invariance from PG;
\item the coframe gravity can be regarded as the gauge-fixed version of the general TG, defined by the Weitzenb\"ock gauge condition $\om^{ij}=0$.
\eitem
Any questions concerning gauge symmetries and the real physical  degrees of freedom of a dynamical system can be most efficiently analyzed in the Dirac Hamiltonian formalism \cite{dirac-1964}.  In the present paper we use this approach to prove the gauge equivalence between the Poincar\'e covariant $\fT$ gravity and its pure coframe form.

This paper is organized as follows. In section \ref{sec2}, we introduce the Lagrangian coframe-connection-multiplier (\ccl\ for short) formulation of the general TG as the subcase of PG, where the condition $R^{ij}=0$ is enforced by a Lagrange multiplier term. Combining  gauge symmetries of the theory with the equations of motion, we show that the only physically relevant dynamical variables are the coframe components $\vth^i$, as expected. In section \ref{sec3}, we use the constrained Hamiltonian approach to make a more detailed analysis of the gauge structure of $\fT$ gravity. Then, in sections \ref{sec4} and \ref{sec5}, we show that gauge symmetries of the theory can be used to identify the associated first class constraints. The result is achieved with the help of the so-called inverse Castellani  algorithm, which largely simplifies  the standard canonical procedure. Finally, choosing the appropriate gauge fixing conditions, the \ccl\ form of $\fT$ gravity is reduced, first to the coframe-connection (\cc), and then to the pure coframe form. In three appendixes, we present some important technical aspects of the analysis and discuss some alternative approaches.

Our conventions are the same as those in Ref.\ \cite{mb.jn-2020}. Latin indices $(i, j,\dots)$ are the local Lorentz indices, greek indices $(\mu,\nu,\dots)$ are the coordinate indices, and both run over $0,1,2,3$; the orthonormal
coframe (tetrad) is $\vth^i=\vth^i{}_\m \rd x^\m$ (1-form), $\vth=\det(\vth^i{}_\m)$, the dual basis (frame) is $e_i= e_j{}^\m\pd_\m$, and $\om^{ij}=\om^{ij}{}_\m \rd x^\m$ is the metric compatible connection (1-form); the metric components in the local Lorentz and coordinate basis are $g_{ij}=(1,-1,-1,-1)$ and $g_{\m\n}= g_{ij}\vth^i{}_\m\vth^j{}_\m$, respectively, the totally antisymmetric symbol $\ve_{ijmn}$ is normalized by $\ve_{0123}=1$, and the square bracket antisymmetrization is defined by  $X^{[iAj]}=(X^{iAj}-X^{jAi})/2$, where $A=\{m...n\}$ is a set of additional indices; wedge products between forms are implicitly understood.

\section{Lorentz invariant form of TG}\label{sec2}
\setcounter{equation}{0}

\subsection{Preliminaries}\label{sub21}

The theories of concern here can be approached via multiple perspectives and representations.  Let us note two in particular.

On the one hand, one can take a \emph{field theory approach} and investigate theories for a dynamical orthonormal \emph{coframe field} $\vth^i$ on spacetime, as was done for example by Itin \cite{Itin}.  A coframe field Lagrangian can have the form $L=L(\vth^i, \rd \vth^i)$.  We stress that such a formulation fundamentally does not logically depend on any conception of parallel transport, connection, curvature or torsion.\footnote{One would, however, need to introduce a connection in order to discuss coupling to sources beyond scalar fields, Maxwell or Yang-Mills gauge fields.} Suppose the dynamical equations determine a unique coframe. Although it is not at all
obligatory, nevertheless the circumstances \emph{naturally} invite one to use the available structure to define \emph{parallel transport} of vectors and tensors along any path as ``keeping the components constant'' in the obtained coframe. This thereby determines a connection of a special type, associated to a global path-independent (hence curvature vanishes) distant parallelism, in other words a \emph{teleparallel} geometry.  In the obtained coframe, the \emph{teleparallel coframe}, the connection $\om^{ij}$ \emph{vanishes}.\footnote{Called the Weitzenb\"ock gauge.}
Now that one has a parallel transport, one can of course represent it in terms of any other local coframe, then the now \emph{non-vanishing} connection $\om^{ij}$ will still have vanishing curvature. No matter what the choice of coframe, the parallel transport is determined by the \emph{torsion tensor}. In this approach teleparallel geometry (represented in general by a coframe and connection) emerges purely from a dynamical coframe field.

On the other hand, one can take a \emph{geometric approach}.  The central concept is \emph{parallel transport} which is determined, with respect to any orthonormal coframe field $\vth^i$, by a set of connection coefficients; these, in turn  determine the torsion and curvature. A geometrically interesting special case, \emph{teleparallel geometry}, has vanishing curvature. The associated parallel transport is path independent. Given a teleparallel geometry, one can start with an orthonormal coframe at any convenient point, parallel transport it along some path (every possible path will give the same result) to every other point, and thereby constructs a global coframe field, the teleparallel coframe,\footnote{ Kopczy\'nski \cite{Kop82} called these frames OT, an acronym for \emph{orthonormal teleparallel}. They are also called Weitzenb\"ock frames.}. which is \emph{unique}, up to a rigid Lorentz transformation. With respect to the teleparallel coframe, the global parallelism is described by the vanishing connection coefficients.  In this approach a preferred global coframe field is determined by a teleparallel geometry.

Geometrically, the natural domain for our considerations is the Riemann-Cartan geometry of spacetime in the Poincar\'e gauge theory, a theory of gravity based on the localization of the Poincar\'e  group of spacetime symmetries (translations and Lorentz rotations) \cite{mb-2002,mb.hehl-2012}. In PG, the basic dynamical variables are the coframe field $\vth^i$ and the metric compatible connection\footnote{For orthonormal coframes, the metric compatibility condition $\nab g_{ij}=0$ implies that $\om^{ij}$ is antisymmetric.} $\om^{ij}$. The corresponding field strengths are the torsion $T^i=\rd\vth^i+\om^i{}_k\vth^k$ and the curvature $R^{ij}=\rd\om^{ij}+\om^i{}_k\om^{kj}$ (2-forms), which satisfy the Bianchi identities
\be
\nab T^i=R^i{}_k \vth^k\, , \qquad \nab R^{ij}=0\,.                  \lab{2.1}
\ee

The general geometric arena of PG can be \emph{a priori} restricted by imposing certain conditions on the field strengths. Thus, the Riemannian geometry of spacetime in GR is defined by the requirement of vanishing torsion,  whereas the Weitzenb{\"o}ck geometry of TG is based on the complementary restriction
\be
R^{ij}(\om)=0\,.                                                     \lab{2.2}
\ee
It ensures, under certain topological assumptions,\footnote{Here, for simplicity, we consider only simply connected spaces or regions. These are parallelizable, admitting globally defined coframe fields. The issues of global topological complications associated with non-simply connected spaces are beyond the scope of the present work.} path independence of the parallel transport. The simplest solution of the condition \eq{2.2} is obtained by choosing $\om^{mn}=0$. In that case, we are left with $\vth^i$ as the only dynamical variable, and the resulting form of TG is known as the coframe gravity. In PG, the \emph{standard} local Lorentz transformation of the (orthonormal) coframe $\vth^i$ is accompanied by an inhomogeneous transformation of the associated connection,
\be
\vth^i\to \L^i{}_m\vth^m\,,\qquad
\om^{ij}\to \L^i{}_m(\L^j{}_n\om^{mn}+\rd\L^{jm})\,,                   \lab{2.3}
\ee
where $\to$ stands for ``is transformed to". Apart from $\om^{mn}=0$, there is another solution of \eq{2.2} given by $\om^{ij}=\L^i{}_m\,\rd\L^{jm}$, which is known as the \emph{pure gauge} connection. It can be related to $\om^{mn}=0$ by a special Lorentz transformation, wherepon the coframe field can be transformed only by \emph{constant} Lorentz transformations.

\subsection{Lagrange multiplier formalism}\label{subsec2.2}

The variation of a Lagrangian in the presence of subsidiary conditions is  mathematically well defined by introducing a suitable Lagrange multiplier term.
Thus, in the framework of PG, the TG Lagrangian can be naturally defined by \cite{mb.in-2000,nester-2018}
\bsubeq\lab{2.4}
\be
L= L_T+\frac{1}{4}\l_{ij}{}^{\m\n}R^{ij}{}_{\m\n}\,,               \lab{2.4a}
\ee
where  $L_T=L_T(\vth^i{}_\m,T^i{}_{\m\n})$ is, in general, any expression invariant under local Poincar\'e transformations, and the $\l R$ term ensures the teleparallelism condition \eq{2.2}\footnote{In the absence of the $\l R$ term, the Lagrangian $L_T$ defines teleparallel gravity only if $\om^{ij}$ is restricted to the pure gauge form. For an application of the Lagrange multiplier formalism to metric--affine gravity, see \cite{hehl-PR95, yo.jp-2003}.}. The standard  PG form of $L_T$ is given as a sum of three independent, quadratic, (parity even) torsion invariants\footnote{More general choices include any function of the three quadratic torsion invariants, and beyond that, higher order torsion invariants.} with arbitrary coefficients,
\be
L_T:=\vth\cL_T\, ,\qquad
     \cL_T=a_0 T^{ijk}(h_1 T_{ijk}+h_2 T_{jik}+h_3 \eta_{ij}V_k)\,, \lab{2.4b}
\ee
\esubeq
where $V_k:=T^m{}_{mk}$. The coframe-connection-multiplier form of the TG  Lagrangian \eq{2.4} is invariant not only under the standard \emph{local Poincar\'e} transformations\footnote{There is one very special case of (2.4b) with certain specific values of $h_1, h_2, h_3$ (the teleparallel equivalent of GR) that has (up to a total differential) an \emph{extra} local Lorentz symmetry acting on the coframe by itself \cite{mb.in-2000}.} (translations with parameters $\xi^\m$, and Lorentz transformations with parameters $\ve^{ij}$),
\bea
&&\d_0\vth^i{}_\m=\ve^i{}_m\vth^m{}_\m-(\pd_\m\xi^\r)\vth^i{}_\r
                                      -\xi^\r\pd_\r\vth^i{}_\m\,,   \nn\\
&&\d_0\om^{ij}{}_\m=-\nab_\m\ve^{ij}-(\pd_\m\xi^\r)\om^{ij}{}_\r
                                      -\xi^\r\pd_\r\om^{ij}\,,      \nn\\
&&\d_0\l_{ij}{}^{\m\n}=\ve_i{}^m\l_{mj}{}^{\m\n}
  + \ve_j{}^m\l_{im}{}^{\m\n}+(\pd_\r\xi^\m)\l_{ij}{}^{\r\n}
  +(\pd_\r\xi^\n)\l_{ij}{}^{\m\r}-\pd_\r(\xi^\r\l_{ij}{}^{\m\n})\,,\qquad
\eea
but also under an extra family of the so-called \emph{lambda transformations}, which will be discussed in the next subsection; see Ref.~\cite{mb.mv-2000}, Eqs. (2.3) and (2.4).

The TG field equations can be conveniently expressed in terms of the covariant momentum associated with the coframe,
\be
H_{ijk}:=\frac{\pd L_T}{\pd T^{ijk}}=\vth\cH_{ijk}\,,\qquad
       \cH_{ijk}=4a_0(h_1 T_{ijk}-h_2 T_{[jk]i}+h_3 \eta_{i[j}V_{k]})\,.
\ee
Indeed, the variation of the Lagrangian \eq{2.4a} with respect to $\vth^i{}_\m,\om^{ij}{}_\m$ and $\l_{ij}{}^{\m\n}$ yields a compact form of the field equations in vacuum (compare to \cite{mb.jn-2020}):
\bsubeq\lab{2.7}
\begin{align}
\cE_i{}^\n:=&-\frac{\d L}{\d\vth^i{}_\n}
 =\nab_\m H_i{}^{\m\n}+T^{mn}{}_i H_{mn}{}^\n-e_i{^\n}L_T=0\,,   \lab{2.7a}\\
\cE_{ij}{}^\n:=&-\frac{\d L}{\d\om^{ij}{}_\n}
              =\nab_\m\l_{ij}{}^{\m\n}+  2H_{[ij]}{^\n}=0\,,     \lab{2.7b}\\
  &~ R^{ij}{}_{\m\n}=0\, .                                       \lab{2.7c}
\end{align}
\esubeq

Let us now focus on some very interesting dynamical properties of these equations.
\bitem
\item[(p1)] The third equation \eq{2.7c} ensures that the geometry of spacetime is teleparallel, which means that the Lorentz connection is a pure gauge (unphysical) variable.
\item[(p2)] The first equation \eq{2.7a}, which is completely determined by $L_T$, does not depend on $\l$.
\item[(p3)] The second equation $\eq{2.7b}$ is the only one that can be used to dynamically determine $\l$. To clarify its content, note that, for $R^{ij}{}_{\m\n}=0$, the covariant divergence of that equation satisfies the 6 Noether identities \eq{A.7},
\be
\nab_\n\cE_{ij}{}^\n +2\cE_{[ij]}\equiv0\,,                                \lab{2.8}
\ee
which do not depend on $\l$. Consequently, the anti-symmetric part of ${\cal E}_{ij}$ contains all the multiplier independent content of ${\cal E}_{ij}{}^\nu$.
\eitem

At this stage, some dynamical aspects of the Lagrange multipliers are still not clear: Equation \eq{2.7b} is an equation with $6\times 4=24$ components, but only $24-6=18$ of them are independent and relevant for the multiplier dynamics. However, that number is not sufficient to determine the $36$ components of $\l_{ij}{}^{\m\n}$.

\subsection{Lambda symmetry}

The worrying aspect of the property (p3) can be better understood by noting that there exists an extra local symmetry acting on the multipliers $\l_{ij}{}^{\m\n}$, which allows one to eliminate some of them as gauge (unphysical) variables \cite{mb.in-2000,nester-2018,mb.mv-2000}. Namely, the Lagrangian \eq{2.4} is, by construction, invariant (up to a divergence) under the local $\l$-transformation
\bsubeq
\be
\d_0\l_{ij}{}^{\m\n}=\nab_\l\t_{ij}{}^{\m\n\l}\, ,            \lab{2.9a}
\ee
where the parameter $\t_{ij}{}^{\m\n\l}$ is completely antisymmetric with respect to  $(\m,\n,\l)$, as well as $(i,j)$. The invariance follows from the Bianchi identity \eq{2.1}$_2$. By expressing $\t_{ij}{}^{\m\n\l}$ in an equivalent form as $\t_{ij}{}^{\m\n\l}=\ve^{\m\n\l\r}w_{ij\r}$, one finds
\be
\d_0\l_{ij}{}^{\m\n}=\ve^{\m\n\l\r}\nab_\l w_{ij\r}\,.
\ee
\esubeq
The number of the local parameters $w_{ij\r}$ is $6\times 4=24$.
However, $w_{ij\l}$ is not uniquely defined, it has its own gauge freedom  (reducibility),
\be
w_{ij\r}\to w_{ij\r}+\nab_\r w_{ij}\,.
\ee
Since the 6 parameters $w_{ij}$ have no influence on $\d_0\l_{ij}{}^{\m\n}$, the effective number of gauge parameters $w_{ij\l}$ is $24-6=18$. They can be used to gauge-fix 18 of the $\l$'s, while the remaining $18$ can be determined by the 18 independent field equations \eq{2.7b}.\footnote{Ref.~[19] notes that there is a possible global topological obstruction to solving this equation for the multiplier for spaces with non-vanishing 3rd cohomology. Some spacetimes of physical interest do have this property, e.g., the Einstein static 3-sphere cosmology.} The above analysis confirms the consistency of the multiplier dynamics.

The only role of Eq.~\eq{2.7c} is to ensure that the spin connection has the pure gauge form. Moreover, from the physical point of view, one can discard Eq.~\eq{2.7b} as it merely determines the (physically uninteresting) Lagrange multiplier, whereas the complete gravitational dynamics is contained in Eq.~\eq{2.7a}. The only dynamical variables in \eq{2.7a} are the coframe field $\vth^i{}_\m$ (16 components) and the pure gauge spin connection $\om^{ij}{}_\m$ (24 components). However, since $\om^{ij}{}_\m$ can be gauge fixed by a suitable choice of the 6 Lorentz parameters $\L^{ik}$, see Eq. \eq{2.3}, one can simply ignore \eq{2.7b} and \eq{2.7c}, and use the 16-component Eq.\ \eq{2.7a} to determine the 16 coframe components.

\subsection{\mb{\fT} gravity}

The dynamical content of the standard quadratic Lagrangian $L_T$ in \eq{2.4b} depends on the values of the coupling constants $h_n$. There is one particularly interesting choice, $(h_1,h_2,h_3)=(1/4,1/2,-1)$, for which the gravitational dynamics, although determined by torsion, is equivalent to GR:
\be
L^{||}:=\vth\mT+\frac{1}{4}\l_{ij}{}^{\m\n}R^{ij}{}_{\m\n}\, ,\qquad
\mT:=\frac{1}{4}a_0T^{ijk}(T_{ijk}+2T_{jik}-4\eta_{ij}V_k)\,.         \lab{2.11}
\ee
This Lagrangian represents the \ccl\ formulation of \tgr.

Inspired by the $f(R)$ extension of GR, one can introduce an analogous extension of \tgr, known as $\fT$ gravity. In order to simplify further analysis, we represent the Lagrangian  $\fT$ in an equivalent form as a Legendre transform of a function $V(\phi)$,
\be
L^f=\vth\cL^f+\frac{1}{4}\l_{ij}{}^{\m\n}R^{ij}{}_{\m\n}\,,
     \qquad \cL^f:=\phi\mT-V(\phi)\,,                                  \lab{2.12}
\ee
where $\phi$ is an auxiliary scalar field. This Lagrangian is invariant under local Poincar\'e transformations, but, as a consequence of the presence of the multiplier and the scalar field, they are slightly modified \cite{mb.mv-2000}. In this representation, the covariant momentum $\cH^f_{ijk}$ is proportional to the \tgr\ form,
\be
\cH^f_{ijk}:=\frac{\pd\cL^f}{\pd T^{ijk}}=\phi\cH_{ijk}\,.
\ee

The previous discussion offers a Lagrangian explanation on how the \ccl\ form of $\fT$ gravity, based on the Lagrangian \eq{2.12}, can be gauge-fixed to obtain the pure coframe form. In what follows, we shall use the Hamiltonian approach to analyze that transition in detail.

\section{Hamiltonian analysis of \mb{\fT} gravity}\label{sec3}
\setcounter{equation}{0}

The canonical analysis of a gauge theory of gravity becomes more efficient by adopting the Dirac-ADM $(1+3)$ formalism \cite{dirac-1964}, which relies on two technical premises: (i) at each point of spacetime, there exists the unit vector $\mb{n}=(n_k)$ normal to the spatial section $\S$ of spacetime, and (ii) any spacetime vector $\mb{V}=(V_k)$ can be projected onto a component $V_\orth:=n^k V_k$ orthogonal to $\S$, and a component $V_\bk:=V_k-n_k V_\orth$ lying in $\S$.

\subsection{Primary constraints}

Starting with the $\fT$ Lagrangian \eq{2.12} and its dynamical variables $(\vth^i{}_\m,\om^{ij}{}_\m,\l_{ij}{}^{\m\n},\phi)$, one can introduce the corresponding canonical momenta $(\pi_i{}^\m,\pi_{ij}{}^\m,\pi^{ij}{}_{\m\n},\p_\phi)$ as
\bsubeq\lab{3.1}
\begin{align}
&\pi_i{}^\m=\frac{\pd L^f}{\pd(\pd_0\vth^i{}_\m)}
            =\phi\,H_i{}^{0\m}\,,&
  &\pi_{ij}{}^\m=\frac{\pd L^f}{\pd(\pd_0\om^{ij}{}_\m)}=\l_{ij}{}^{0\m}\,,&\\
&\pi^{ij}{}_{\m\n}=\frac{\pd L^f}{\pd(\pd_0\l_{ij}{}^{\m\n})}=0\,,&
  &\pi_\phi=\frac{\pd\cL^f}{\pd(\pd_0\phi)}=0\,.&
\end{align}
\esubeq
The absence of time derivatives of $(\vth^i{}_0,\om^{ij}{}_0,\l_{ij}{}^{\m\n},\phi)$ implies the following (sure) primary constraints:
\bsubeq\lab{3.2}
\begin{align}
&\phi_i{^0}:=\pi_i{^0}\approx 0\,,&
  &\phi_{ij}{^0}:=\pi_{ij}{^0}\approx 0\,, &                      \lab{3.2a}\\
&\phi^{ij}{}_{\a\b}:=\pi^{ij}{}_{\a\b}\approx 0\,,&
  &\phi^{ij}{}_{0\b}:=\pi^{ij}{}_{0\b}\approx 0\,,&               \lab{3.2b}\\
&&&\phi_{ij}{^\a}:=\pi_{ij}{^\a}-\l_{ij}{}^{0\a}\approx 0\,,&     \lab{3.2c}\\
&&&\pi_\phi\approx 0\,.&                                          \lab{3.2d}
\end{align}
\esubeq

The specific form of \mT\ produces some ``extra" constraints, which can be found by rewriting the relation $\pi_i{}^\a=\phi H_i{}^{0\a}$ as
\be
\hpi_{i\bk}=\phi J\cH_{i\orth\bk}\, ,                              \lab{3.3}
\ee
where  $\hpi_{i\bk}:=\pi_i{}^\b\vth_{k\b}$ are the ``parallel" canonical momenta.
Indeed, by analyzing this 12-component equation, one finds that the components which are independent of the ``velocities" $T_{i\orth\bk}$ can be combined to obtain 6 new constraints \cite{mb.in-2000,mb.jn-2020},
\begin{align}
&C_{ik}=\hpi_{i\bk}-\hpi_{k\bi}+a_0\phi B_{ik}\approx 0\,,        \lab{3.4}\\
&B_{ik}:=\nab_\a B^{0\a}_{ij}\,,\qquad
         B^{0\a}_{ij}:=\ve^{0\a\b\g}_{ikmn}\vth^m{}_\b\vth^n{}_\g\,.\nn
\end{align}
The remaining 6 components can be solved for the 6 velocities $\ir{T}T_{\bi\orth\bk}$ and $T^\bm{}_{\orth\bm}$.

\subsection{Hamiltonians}

Having found all the primary constraints, we now introduce the  canonical Hamiltonian,
\be
H_c=\pi_i{^\a}\pd_0\vth^i{_\a}+\frac{1}{2}\pi_{ij}{}^\a\pd_0\om^{ij}{}_\a
                              +\pi_\phi\pd_0\phi-L^f\,.
\ee
Since $L^f$ is linear in $\pd_0\om^{ij}{}_\a$, as follows from
$R^{ij}{}_{0\a}\equiv\pd_0 \om^{ij}{}_\a-\nab_\a\om^{ij}{}_0$, one can use $\phi_{ij}{}^\a\approx 0$ to obtain
\be
\pi_{ij}{^\a}\pd_0\om^{ij}{}_\a\approx
     \l_{ij}{}^{0\a}(R^{ij}{}_{0\a}+\nab_\a\om^{ij}{}_0)\,.\nn
\ee
Then, after eliminating the coframe velocities with the help of the relations defining $T^i{}_{0\a}$,
\be
T^i{}_{0\a}\equiv \pd_0\vth^i{_\a}+\om^i{}_{k0}\vth^k{_\a}-\nab_\a\vth^i{_0}
                 = N T^i{}_{\orth\a}+N^\b T^i{}_{\b\a}\,,            \nn
\ee
where $N$ and $N^\a$ are the lapse and shift functions, respectively, and using  the (1+3) decomposition $\pd_0\phi=N\pd_\orth\phi+N^\b \pd_\b\phi$, the canonical Hamiltonian takes the Dirac-ADM form
\bsubeq\lab{3.6}
\be
H_c=N\cH_\orth+N^\a\cH_\a-\frac{1}{2}\om^{ij}{_0}\cH_{ij}
       -\frac{1}{4}\l_{ij}{}^{\a\b}R^{ij}{}_{\a\b}+\pd_\a D^\a\,,
\ee
where
\bea
&&\cH_\orth:=\hpi_i{^\bm}T^i{}_{\orth\bm}-J\cL^f
                     -n^k\nab_\b\pi_k{^\b}+\hu_\phi\pi_\phi\,,  \nn\\
&&\cH_\a:=\pi_k{^\b}T^k{}_{\a\b}-\vth^k{_\a}\nab_\b\pi_k{^\b}
                                              +\pi_\phi\pd_\a\phi\,, \nn\\
&&\cH_{ij}:=\hpi_{i\bk}-\hpi_{k\bi}+\nab_\a\pi_{ij}{}^\a\,,          \nn\\
&&D^\a:=\vth^k{_0}\pi_k{^\a}+\frac{1}{2}\om^{ij}{_0}\p_{ij}{^\a}\,.\lab{3.6b}
\eea
\esubeq
Following the spirit of the Dirac algorithm, the $\pi_\phi$ term in $\cH_\orth$ is written in the form $\hu_\phi\pi_\phi$, where $\hu_\phi$ is an independent ``velocity" multiplier, defined by $\hu_\phi:=\pd_\orth\phi$. The canonical Hamiltonian is linear in the unphysical variables $\vth^i{}_0,\om^{ij}{}_0$ and $\l_{ij}{}^{\a\b}$. The lapse Hamiltonian $\cH_\orth$ is the only dynamical component of $H_c$, as it depends on the Lagrangian. Using \eq{3.3} to eliminate the velocities $\pd_0\vth^i{}_\a$ from $\cH_\orth$, one obtains \cite{mb.jn-2020}
\begin{align}
\cH_\orth=&\,\frac{1}{2a_0\phi}P^2-J(\phi\bar\mT-V)
                      -n_i\nab_\a\pi_i{}^\a+\hu_\phi\pi_\phi\,,      \nn\\
P^2:=&\,\frac{1}{2J}\Big[\hpi_{\bm\bn}\hpi^{\bm\bn}
                       -\frac{1}{2}(\hpi^\bm{}_\bm)^2\Big]\,,        \nn\\
\bar\mT:=&\,\frac{1}{4}a_0\Big(T^{\bi\bm\bn}T_{\bi\bm\bn}
    +2T^{\bm\bi\bn}T_{\bi\bn\bm}
                     -4T^{\bm}{}_{\bm\bk}T_{\bn}{}^{\bn\bk}\Big)\,.
\end{align}

The general Hamiltonian dynamics is determined by the total Hamiltonian
\begin{align}
H_T=&\,H_c+u^i{_0}\pi_i{^0}+\frac{1}{2}u^{ij}{_0}\pi_{ij}{^0}
        +\frac{1}{2}u^{ij}{_\a}\phi_{ij}{^\a}                       \nn\\
&+\frac{1}{2}u_{ij}{}^{0\b}\pi^{ij}{}_{0\b}
  +\frac{1}{4}u_{ij}{}^{\a\b}\pi^{ij}{}_{\a\b}+(v\cdot C)\,,       \lab{3.8}
\end{align}
where $v\cdot C:=\frac{1}{2}v^{mn}C_{mn}$. In Ref.~\cite{mb.jn-2020}, the term $\pi_\phi\pd_0\phi$ is relocated from $H_c$ to $H_T$, where it appears in the form $u_\phi\pi_\phi$.

\subsection{Preservation of the primary constraints}

The preservation of the primary constraints \eq{3.2a} yields
\bsubeq\lab{3.9}
\bea
-\pd_0\pi_i{^0}:
    &&\chi_i:=\cH_i= n_i\cH_\orth+e_\bi{}^\a\cH_\a\approx 0\,,     \lab{3.9a}\\
\pd_0\pi_{ij}{^0}: &&\chi_{ij}:=\cH_{ij}\approx 0\,.               \lab{3.9b}
\eea
\esubeq
Similarly, the preservation of \eq{3.2b} is given by
\bsubeq
\bea
\pd_0\pi^{ij}{}_{\a\b}:&&\chi^{ij}{}_{\a\b}
                         :=R^{ij}{}_{\a\b}\approx 0\, ,             \lab{3.10a}\\
\pd_0\pi^{ij}{}_{0\b}:&&\chi^{ij}{}_{0\b}
                    :=u^{ij}{_\b}\approx 0\,.                       \lab{3.10b}
\eea
\esubeq
Since the equation of motion for $\om^{ij}{_\a}$ has the form
\be
\pd_0\om^{ij}{_\a}=\nab_\a\om^{ij}{_0}+u^{ij}{_\a}\quad\Lra\quad
   R^{ij}{}_{0\a}=u^{ij}{_\a}\approx 0\,,                            \lab{3.11}
\ee
it follows that all components of the curvature tensor weakly vanish, as expected.

The preservation of the primary constraints \eq{3.2c} takes the general form
\bea
\pd_0\phi_{ij}{^\a}:\quad
u_{ij}{}^{0\a}=\{\phi_{ij}{^\a},H_c+(v\cdot C)\}\,.                  \lab{3.12}
\eea
Similarly, the preservation of $\pi_\phi$ yields
\bsubeq\lab{3.13}
\be
\pd_0\pi_\phi:\quad \chi_\phi
                       :=NF_\phi-\frac{1}{2}v^{mn}F_{mn}\approx 0\,, \lab{3.13a}
\ee
where
\bea
&&F_\phi:=\{\pi_\phi,\cH_\orth\}
         =\frac{1}{2a_0\phi^2}P^2+J(\bar\mT-\pd_\phi V)\,,             \nn\\
&&F_{mn}:=-\{\pi_\phi,C_{mn}\}=a_0 B_{mn}\,.
\eea
\esubeq
Finally, as for the preservation of the extra primary constraints $C_{ij}$,  we have
\be
\pd_0 C_{ij}:\quad\chi_{ij}:=G_{ij}{}^k(\pd_k\phi)\d\approx 0\,,  \lab{3.14}
\ee
where the  (complicated, multiplier dependent) coefficients $G_{ij}{}^k$ can be found in Ref.~\cite{mb.jn-2020}, Eq.~(3.17), modified by $\pd\to\nab$.

One can summarize the present situation as follows: the preservation of the primary constraints \eq{3.2} yields the secondary constraints \eq{3.9}, \eq{3.10a}, \eq{3.13a} and \eq{3.14}, plus the conditions \eq{3.10b} and \eq{3.12} on the multipliers. Hence, $\fT$ gravity is described by the total Hamiltonian \eq{3.8}, with the set of constraints and canonical multipliers (the fixed ones are marked by a bar), displayed in Table 1.

\begin{center}
\doublerulesep 1.8pt
\begin{tabular}{l|llll}
\multicolumn{5}{c}{Table 1. Constraints and multipliers,
                                               the present status} \\
                                                          \hline\hline
\rule[-1pt]{0pt}{15pt}
primary &~$\pi_i{}^0\,,\pi_{ij}{}^0$ & $\pi^{ij}{}_{\a\b}$ & $\pi_\phi, C_{mn}$
                               & $\pi^{mn}{}_{0\b}\,,\phi_{ij}{}^\a$  \\
\rule[-1pt]{0pt}{15pt}
secondary
    &~$\cH_\orth\,,\cH_\a,\cH_{ij}$  & $R^{ij}{}_{\a\b}$ & $\chi_\phi,\chi_{mn}$ & \\[2pt]
\rule[-1pt]{0pt}{15pt}
multipliers &~$u^i{}_0\,,u^{ij}{}_0$ & $u_{ij}{}^{\a\b}$ & $\hu_\phi,v_{mn}$
                   & $\bu_{ij}{}^{0\b}\,,\bu^{ij}{}_\a$           \\[2pt]
                                                         \hline\hline
\end{tabular}
\end{center}
The vanishing of $\bu^{ij}{}_\a$ is the canonical counterpart of the Lagrangian relation $R^{ij}{}_{0\a}=0$.

\subsection{Preliminary Dirac brackets}

Further analysis can be simplified by noting that the primary constraints $(\pi^{mn}{}_{0\b},\phi_{ij}{}^\a)$, which are second class, can be used to introduce the preliminary \emph{Dirac brackets} (DBs). Then, one can use these constraints as strong equalities and eliminate the pair of the canonically conjugate variables $(\l_{ij}{}^{0\a},\pi^{mn}{}_{0\b})$ from the theory,  $\l_{ij}{}^{0\a}=\pi_{ij}{}^\a$ and $\pi^{mn}{}_{0\b}=0$. After that, DBs in the remaining phase space reduce to Poisson brackets (PBs).\footnote{Ater eliminating $\l_{ij}{}^{0\a}$, any variable from the reduced phase space has a vanishing PB with $\pi^{mn}{}_{0\b}$. Then, by construction, any DB reduces just to the PB form.}

In such a reduced phase space, denoted by $R_1$, the terms  $\bu_{ij}{}^{0\a}\pi^{ij}{}_{0\a}$ and $\bu^{ij}{}_\a\phi_{ij}{}^\a$ in $H_T$ strongly vanish, so that
\be
H_T(R_1)=H_c+u^i{_0}\pi_i{^0}+\frac{1}{2}u^{ij}{_0}\pi_{ij}{^0}
  +\frac{1}{4}u_{ij}{}^{\a\b}\pi^{ij}{}_{\a\b}+(v\cdot C)\,.       \lab{3.15}
\ee
The constraints characterizing the phase space $R_1$ are given in Table 2.
\begin{center}
\doublerulesep 1.8pt
\begin{tabular}{l|c|l}
\multicolumn{3}{c}{Table 2. Constraints in $R_1$}     \\
                                                          \hline\hline
   & sure constraints &  extra   \\
\hline
\rule[-1pt]{0pt}{15pt}
primary
   & $\pi_i{}^0\,,\pi_{ij}{}^0,\hfill \pi^{ij}{}_{\a\b}$ &
                                                 $\pi_\phi,C_{mn}$ \\
\rule[-1pt]{0pt}{15pt}
secondary
   & $\cH_\orth\,,\cH_\a,\cH_{ij},\quad R^{ij}{}_{\a\b}$ &
                                                 $\chi_\phi,\chi_{mn}$ \\[2pt]
                                                         \hline\hline
\end{tabular}
\end{center}
In $R_1$, the relation $R^{ij}{}_{0\b}=0$ is a dynamical consequence of  $\bu^{ij}{}_\b=0$.

\subsection{Preservation of the secondary constraints}

Now, we are going to examine the preservation of the sure secondary constraints in $R_1$.

Regarding the preservation of $R^{ij}{}_{\a\b}$, one can use the equation of motion \eq{3.11} for $\om^{ij}{}_\a$ to obtain
\be
\nab_0 R^{ij}{}_{\a\b}=\nab_\a\bu^{ij}{}_\b-\nab_\b\bu^{ij}{}_\a\,,  \lab{3.16}
\ee
which implies that the preservation of $R^{ij}{}_{\a\b}$ is automatically satisfied. Since $R^{ij}{}_{0\b}\approx\bu^{ij}{}_\b$, the relation \eq{3.16} can be interpreted as the weak version of the second Bianchi identity \eq{2.1}$_2$.

As follows from the analysis of the translational and local Lorentz symmetry in Appendix \ref{appB} and subsection \ref{sub51}, respectively, the Hamiltonian constraints $(\cH_\orth,\cH_\a)$ and $\cH_{ij}$ must be first class. Thus, the secondary constraints $(\cH_\orth,\cH_\a,\cH_{ij},R^{ij}{}_{\a\b})$ are preserved independently of the status of any other constraint in $R_1$.

We defer discussion of preserving the extra constraints $(\chi_\phi,\chi_{ij})$ until the end of section \ref{sec5}.

\section{Gauge fixing the local lambda symmetry}\label{sec4}
\setcounter{equation}{0}

In this section, we discuss how  the \ccl\ formulation of $\fT$ gravity can be reduced to the \cc\ form, by gauge fixing the local lambda symmetry.

\subsection{The inverse Castellani algorithm}\label{sub41}

In 1918 Noether presented two theorems regarding symmetry in dynamical systems. Briefly, they stated that for each global or local Lagrangian symmetry there is a conserved current or differential identity, and conversely, see for instance \cite{chen-2015}.  The Hamiltonian counterpart of these theorems is richer: the conserved current is the generator of the symmetry, furthermore local symmetries are associated with first class constraints. For local symmetries, Castellani \cite{lc-1982} showed in detail how to construct the gauge generator from the first class constraints.  As we shall demonstrate in the $f(\mT)$ case, the converse of the Castellani construction can be quite useful: if one knows that the Lagrangian has certain local symmetries, one can infer that the Hamiltonian formulation has corresponding gauge generators containing canonical constraints, and these constraints are necessarily first class. This can be quite helpful in the process of identifying and classifying the constraints in a complicated system.

The gauge fixing procedure of a local symmetry is closely related to the form of the associated first class constraints. The standard canonical procedure for identifying first class constraints requires one to complete the Hamiltonian constraint analysis, which includes finding all the constraints and calculating their PB algebra \cite{dirac-1964}. In $\fT$ gravity, such a program would be extremely complicated, see, for instance, Ref.\ \cite{mb.jn-2020}. Fortunately, there is a simpler and more practical approach to the problem, described by the following set of instructions:
\bitem
\item[(i1)] one starts from the known canonical generator of a local symmetry in \tgr;
\item[(i2)] then, by the ``guess and check" strategy, one tries to find its correct form in $\fT$ gravity;
\item[(i3)] once the correct generator is found, one can immediately identify the associated first class constraints in $\fT$ gravity.
\eitem

Let us now apply this approach to the local lambda symmetry. In \tgr, the corresponding canonical gauge generator has the form \cite{mb.mv-2000}
\begin{align}
G&\,=G_A+G_B\, ,                                                    \nn\\
G_A&:=-\frac{1}{4}\dot\t_{ij}{}^{\a\b}\pi^{ij}{}_{\a\b}
       +\frac{1}{4}\t_{ij}{}^{\a\b}S^{ij}{}_{\a\b}\,,\quad
  S^{ij}{}_{\a\b}:=-4R^{ij}{}_{\a\b}+2\om^{[i}{}_{k0}\pi^{j]k}{}_{\a\b}\,,\nn\\
G_B&:=-\frac{1}{4}\t_{ij}{}^{\a\b\g}\nab_\a\pi^{ij}{}_{\b\g}\,. \lab{4.1}
\end{align}
The result is obtained using the systematic Castellani algorithm \cite{lc-1982}, based on the fact that the constraints ($\pi^{ij}{}_{\a\b},R^{ij}{}_{\a\b}$) are first class.

Since the local lambda symmetry in $\fT$ gravity is generated by the same mechanism as in \tgr, stemming from the presence of the $\l R$ terms in their Lagrangians, it is not surprising that it has the same form in both theories. Thus, one can infer that $G$ is also the correct generator in the framework of $\fT$ gravity,\footnote{The statement can be explicitly checked using the transformation rule $\d_0\vphi_A:=\{\vphi_A,G\}$. A direct calculation yields the result which is in agreement with the corresponding Lagrangian formula \eq{2.9a}, provided $\t_{ij}{}^{\a\b}$ is identified with $\t_{ij}{}^{\a\b 0}$ \cite{mb.mv-2000}. The conclusion holds also for the momentum variables $\pi_{ij}{}^\m$.} reduced to the phase space $R_1$.\footnote{The whole gauge fixing procedure can be easily extended to the original, unrestricted phase space described in Table 1, relying on the results of Ref.~\cite{mb.mv-2000}.} Hence:
\bitem
\item[(a1)] Since $G$ is the correct gauge generator for $\fT$ gravity, the constraints  $(\pi^{ij}{}_{\a\b},S^{ij}{}_{\a\b})$, or equivalently $(\pi^{ij}{}_{\a\b},R^{ij}{}_{\a\b})$, are first class, regardless of the existence of any other constraint.
\eitem
Indeed, if $(\pi^{ij}{}_{\a\b},R^{ij}{}_{\a\b})$ were not first class, the form of the generator \eq{4.1} would be in contradiction to the Castellani algorithm. Hence, ($\pi^{ij}{}_{\a\b},R^{ij}{}_{\a\b}$) must be first class.

The procedure just described can be naturally called the \emph{``inverse" Castellani algorithm:} starting with the generator of local lambda symmetry, one concludes that the constraints multiplying the parameters $\dot\ve$ and $\ve$,  are first class. In fact, a more general treatment of the same idea has been used in PG \cite{mb.mv-1987,in-1992}, not only to identify first class constraints, but also to determine their PB algebra.

\subsection{The coframe-connection form of \mb{\fT} gravity}

Having found that $\pi_{ij}{}^{\a\b}$ are first class, we choose  $\l_{ij}{}^{\a\b}\approx 0$ as the associated gauge conditions. Treating these conditions as any other constraint, we find that their preservation takes the form $\pd_0\l_{ij}{}^{\a\b}=u_{ij}{}^{\a\b}=0$. Hence, the momentum $\pi_{ij}{}^{\a\b}$ disappears from $H_T(R_1)$, remaining dynamically decoupled from the rest of $R_1$. Such a status of the pair  $(\l_{ij}{}^{\a\b},\pi^{mn}{}_{\g\d})$ can be equivalently described by constructing the corresponding DBs and eliminating these variables from $R_1$ via the strong equalities $\l_{ij}{}^{\a\b}=0,$ $\pi^{ij}{}_{\a\b}=0$. The DBs among the remaining variables take the PB form, and the new, reduced phase space $R_2$ is described in Table 3.
\begin{center}
\doublerulesep 1.8pt
\begin{tabular}{l|c|l}
\multicolumn{3}{c}{Table 3. Constraints in $R_2$}     \\
                                                      \hline\hline
  & sure constraints & extra                          \\
\hline
\rule[-1pt]{0pt}{15pt}
primary &~$\pi_i{}^0\,,\pi_{ij}{}^0\hfill -\quad$  & $\pi_\phi,C_{mn}$ \\
\rule[-1pt]{0pt}{15pt}
secondary &~$\cH_\orth\,,\cH_\a,\cH_{ij},\quad R^{ij}{}_{\a\b}$
                                           & $\chi_\phi,\chi_{mn}$ \\[2pt]
                                                        \hline\hline
\end{tabular}
\end{center}

Taking into account that $\l_{ij}{}^{\a\b}$ strongly vanishes, the total Hamiltonian now reads
\begin{align}
H_T(R_2)=\,
  &N\cH_\orth+N^\a\cH_\a-\frac{1}{2}\om^{ij}{_0}\cH_{ij}+\pd_\a D^\a \nn\\
      &+u^i{_0}\pi_i{^0}
               +\frac{1}{2}u^{ij}{_0}\pi_{ij}{^0}+(v\cdot C)\,.    \lab{4.2}
\end{align}
Here, the Hamiltonian constraints $(\cH_\orth,\cH_\a,\cH_{ij})$ contain the nontrivial covariant derivatives $\nab_\a$, in contrast to the situation in pure coframe gravity.
\bitem
\item[(a2)] By fixing the local lambda symmetry, the present canonical formulation corresponds to the \cc\ form of $\fT$ gravity, defined by the Lagrangian \eq{2.12} with $\l_{ij}{}^{\m\n}=0$ and $R^{ij}{}_{\m\n}=0$, that is with
    \be
    L^f=\vth[\phi\mT-V(\phi)]\,,
    \ee
    where $\om^{ij}{}_\m$ is a pure gauge connection.
\eitem

\section{Gauge fixing the local Lorentz symmetry}\label{sec5}
\setcounter{equation}{0}

\subsection{First class constraints in the Lorentz sector}\label{sub51}

As we mentioned in section \ref{sub21}, the transition from the pure gauge connection to $\om^{ij}{}_\m=0$ can be understood as a reduction of the local Lorentz symmetry to its rigid form. To examine that conclusion in the Hamiltonian formalism, we start from the canonical generator of the standard local Lorentz transformations in \tgr\ \cite{mb.mv-2000}. After reducing it to the phase space $\bR_2$ of \tgr, it takes the form
\be
G_L:=-\frac{1}{2}\dot\ve^{ij}\pi_{ij}{}^0-\frac{1}{2}\ve^{ij}S_{ij}\,,\qquad
  S_{ij}:=-\cH_{ij}+2\vth_{[i0}\pi_{j]}{}^0
                          +2\om^k{}_{[i0}\pi_{kj]}{}^0\,.           \lab{5.1}
\ee
The above generator produces the correct Lorentz transformations of the fields $(\vth^k{}_\m,\om^{ij}{}_\n)$ and their conjugate momenta, in accordance with the standard PG rules.

However, the phase space $R_2$ of $\fT$ gravity contains two additional variables, $\phi$ and $\pi_\phi$, on which the generator \eq{5.1} acts trivially,
\be
\d_0\phi=\{\phi,G_L\}=0\, ,\qquad \d_0\pi_\phi=\{\pi_\phi,G_L\}=0\,.
\ee
How can this result help us in a search for the true Lorentz generator in $\fT$ gravity? Recalling that $\phi$ is a scalar field, it follows that the first relation is, in fact, quite correct. Similarly, since $\pi_\phi\approx 0$, the second relation is also correct. Thus, the Lorentz gauge generator \eq{5.1} is also the proper generator of Lorentz transformations in $\fT$ gravity.
\bitem
\item[(b1)] Hence, as a consequence of the inverse Castellani algorithm, the constraints $(\p_{ij}{}^0,S_{ij})$, or equivalently $(\pi_{ij}{}^0,\cH_{ij})$, are necessarily first class.
\eitem
The conclusion is based on the fact that $\pi_j{}^0$ is first class, as shown in Appendix \ref{appB}.

\subsection{The coframe form of \mb{\fT} gravity}

The reduction of $\fT$ gravity to the coframe form requires a suitable mechanism for breaking the local Lorentz symmetry. That mechanism can be understood as follows.

First, starting from the first class constraint $\pi_{ij}{}^0$, we impose the gauge condition
\bsubeq
\be
\om^{ij}{}_0\approx 0\,.                                           \lab{5.3a}
\ee
Its preservation takes the form $\pd_0\om^{ij}{}_0=u^{ij}{}_0=0$. Then, for consistency, local Lorentz symmetry has to be restricted to a subclass, defined by
\be
\d_0\om^{ij}{}_0=\{\om^{ij}{}_0,G_L\}=\pd_0\ve^{ij}=0\,.
\ee
\esubeq
The pair $(\om^{ij}{}_0,\pi_{ij}{}^0)$ can be eliminated from $R_2$ by constructing the corresponding DBs.

And second, the first class nature of $S_{ij}$ motivates us to impose another gauge condition
\bsubeq
\be
\om^{ij}{}_\a\approx 0\, .                                        \lab{5.4a}
\ee
It is automatically preserved, as follows from the relation \eq{3.11}, and its consistency with local Lorentz symmetry takes the form
\be
\d_0\om^{ij}{}_\a=\{\om^{ij}{}_\a,G_L\}\approx -\pd_\a\ve^{ij}=0\,.
\ee
\esubeq

\bitem
\item[(b2)] The gauge fixing conditions \eq{5.3a} and \eq{5.4a} imply that the local Lorentz invariance is reduced to its rigid form.
\eitem
In the resulting phase space $R_3$, the total Hamiltonian reads
\be
H_T(R_3)=
  N\cH_\orth+N^\a\cH_\a+u^i{_0}\pi_i{^0}+(v\cdot C)\,,             \lab{5.5}
\ee
It represents the canonical description of the coframe formulation of $\fT$ gravity, defined by the Lagrangian \eq{2.12} with $\l_{ij}{}^{\m\n}=0$ and $\om^{ij}{}_\r=0$.

\bitem
\item[$\bullet$] All the previous considerations imply our \emph{main result:}\\
The \ccl, the \cc, and the pure coframe forms of $\fT$ gravity, are just three gauge equivalent versions of the same theory.
\eitem

At this point, we add a brief remark on the preservation of the extra constraints $(\chi_\phi,\chi_{ij})$. The rather complicated details of how one uses $\chi_\phi$ and $\chi_{ij}$ to find one further tertiary constraint and determine the multipliers $u_\phi\equiv\pd_0\phi$ and $v_{mn}$, are given in our paper \cite{mb.jn-2020}. However, the results obtained here do not depend on these details.

\section{Concluding remarks}
\setcounter{equation}{0}

In this work, we analyzed the gauge structure of $\fT$ gravity using the Dirac Hamiltonian approach, in which, as explicitly shown by Castellani \cite{lc-1982}, gauge symmetries are generated by first class constraints. However, the standard procedure for identifying these constraints requires a large number of the PB calculations. To avoid these complications, we introduced another, much simpler but equally reliable method, suitably named the inverse Castellani algorithm, see section \ref{sec4}. After having found all the first class constraints associated to the lambda, the Lorentz, and the translational gauge symmetries, we applied the standard canonical gauge fixing procedure to show that the \ccl, the \cc, and the pure coframe formulations of $\fT$ gravity are gauge equivalent. As a direct consequence, the three formulations have the same number of the physical degrees of freedom.

There is another interesting consequence  that implicitly follows from our analysis. Namely, since the set of first class constraints found through the inverse Castellani algorithm does not depend on the existence of any other constraint in the theory, the present proof of gauge equivalence can be  straightforwardly extended to any Poincar\'e gauge invariant TG. In particular, it holds for $\fT$ gravity, or for its extension in which $\mT$ is replaced by a more general torsion invariant, such as the one defined in Eq.~\eq{2.4b}, or its subcase known as New general relativity (NGR) \cite{hs-1979}.\footnote{A Hamiltonian analysis of NGR as a pure coframe theory was given in Ref.~\cite{cheng-1988}.} With the known set of first class constraints, the full Hamiltonian analysis becomes notably simpler.

In our approach, the \cc\ form of TG is obtained from its \ccl\ version by fixing the lambda gauge symmetry. As a result, the original Lorentz gauge symmetry is left untouched, it continues to be a valid symmetry of the \cc\ formulation. The only remnant of the original formulation is an extra geometric restriction, the teleparallelism condition $R^{ij}=0$, which is independent of, but fully compatible with the Lorentz gauge symmetry.
As a consequence, the flat spin connection $\om^{ij}$ can be varied \emph{off shell} as an ordinary Riemann--Cartan connection. This approach yields the standard form of the Noether differential identity \eq{A.7}, which relates the second field equation of $\fT$ gravity \eq{2.7b} with the first one, \eq{2.7a}; see also \cite{nester-2018}. The result is quite general, it follows only from the gauge structure of the theory and offers a clear description of the dynamical role of the second field equation; compare to \cite{golovnev-2017,krssak-2018}.

The geometric structure of the \cc\ form of  TG is quite naturally described as a subcase of the Riemann-Cartan geometry. Still, there are opinions, see for instance \cite{krssak-2015,bejarano-2019}, that the covariant TG can be more conveniently defined by replacing the vanishing spin connection of the pure coframe TG with its Lorentz transform  $\om^{ij}(\L)=\L^i{}_m \rd\L^{jm}$. In this coframe-Lorentz formalism the basic dynamical variables are ($\vth^i,\L^i{}_m)$, in contrast to the genuine \emph{geometric} choice $(\vth^i,\om^{ij})$. At present, it is not yet clear to what extent the coframe-Lorentz formalism will become a useful alternative to the more geometric \cc\ approach; for more details, see Appendix \ref{appC}.

Much of what we did in the present work would easily extend to $D>4$ dimensions. We also wish to remark that here, we considered only metric compatible connections. Symmetric teleparallel connections with nonmetricity are also of interest \cite{nester.yo-1999}, but are not covered in the present work.

\appendix
\section{Noether identities}\label{appA}
\setcounter{equation}{0}

Consider a general TG Lagrangian \eq{2.4a}, and introduce the notation
\bea
&& H_i{}^{\m\n}:=\frac{\pd L}{\pd T^i{}_{\m\n}}\,, \qquad
   H_{ij}{}^{\m\n}:=\frac{\pd L}{\pd R^{ij}{}_{\m\n}}\,,\qquad
   E_i{}^\m:=\frac{\pd L}{\pd\vth^i{}_\m}\,,                         \nn\\
&&\cE_i{}^\n:=\nab_\m H_i{}^{\m\n}-E_i{}^\n\,,\qquad
  \cE_{ij}{}^\n:=\nab_\m H_{ij}{}^{\m\n}+H_{ij}{}^\n-H_{ji}{}^\n\,.\lab{A.1}
\eea
Then, using the formulas
\begin{align}
   \d T^i{}_{\m\n}&=(\nab_\m\d\vth^i{}_\n+\d\om^i{}_{k\m}\vth^k{}_\n)
                                           -(\m\lra\n)\,,            \nn\\
\d R^{ij}{}_{\m\n}&=\nab_\m\d\om^{ij}{}_\n-(\m\lra\n)\,,
\end{align}
one can calculate the general variation of the Lagrangian, $\d L=\d_1 L+\d_2 L$:
\bsubeq\lab{A.3}
\begin{align}
\d_1L:=&\,\,\frac{1}{2}\d T^i{}_{\m\n}H_i{}^{\m\n}
       +\frac{1}{4}\d R^{ij}{}_{\m\n}H_{ij}{}^{\m\n}
       +\d\vth^i{}_\m E_i{}^\m                                        \nn\\
   =&\,\,\nab_\m\Big(\d\vth^i{}_\n H_i{}^{\m\n}
              +\frac{1}{2}\d\om^{ij}{}_\n H_{ij}{}^{\m\n}\Big)
  -\d\vth^i{}_\n\,\cE_i{}^\n -\frac{1}{2}\d\om^{ij}{}_\n\,\cE_{ij}{}^\n\,,\\
\d_2 L:=&\,\,\frac{1}{4}(\d\l_{ij}{}^{\m\n})R^{ij}{}_{\m\n}\,.
\end{align}
\esubeq

Let us now assume the invariance of $L$ under the local Lorentz transformations,
\be
\d\vth^i{}_\m=\ve^i{}_k\vth^k{}_\m\,,\qquad\d\om^{ij}{}_\m=-\nab_\m\ve^{ij}\,,
\qquad\d\l_{ij}{}^{\m\n}=\ve_i{}^k\l_{kj}{}^{\m\n}-\ve_j{}^k\l_{ki}{}^{\m\n}\,,
\ee
as well as the validity of the teleparallelism condition $R^{ij}{}_{\m\n}=0$. Then, $\d L$ takes the form
\begin{align}
\d_\ve L&=\nab_\m\Big[-\ve^{ij} H_{ij}{}^\m
               -\frac{1}{2}\ve^{ij}\nab_\r H_{ij}{}^{\r\m}\Big]
 -\ve^i{}_k\vth^k{_\m}\cE_i{}^\m+\frac{1}{2}(\nab_\m\ve^{ij})\cE_{ij}{}^\m\nn\\
&=\frac{1}{2}(\nab_\m\ve^{ij})
  \Big(\cE_{ij}{}^\m-\nab_\r H_{ij}{}^{\r\m}-2H_{ij}{}^\m\Big)
  -\frac{1}{2}\ve^{ij}\Big[\nab_\m\big(\ub{\nab_\r H_{ij}{}^{\r\m}+2H_{ij}{}^\m}_{\cE_{ij}{}^\m}\big)+2\cE_{ij}\Big]\,.
\end{align}
Since $\d_\ve L=0$, the coefficients of the $\ve^{ij}$ and  $\nab_\m\ve^{ij}$ terms vanish separately. The vanishing of the $\nab\ve$ term defines the Noether current
\be
J_{ij}{}{}^\m:=\nab_\r H_{ij}{}^{\r\m}+2H_{[ij]}{}^\m\equiv\cE_{ij}{}^\m\,,
\ee
whereas the vanishing of the $\ve$ term yields the Noether differential identity
\be
\nab_\m\cE_{ij}{}^\m+2\cE_{[ij]}\equiv0\,.                                \lab{A.7}
\ee
These identities are obtained without assuming the Euler-Lagrange field equations, i.e., the vanishing of $\cE_i{}^\m$ and $\cE_{ij}{}^\m$. Equivalent results can be found in earlier works, e.g. Refs.~\cite{nester-2018,hehl-PR95,yo.jp-2003}.

\section{Local translations and first class constraints}\label{appB}
\setcounter{equation}{0}

In this appendix, we use the inverse Castellani algorithm to identify the first class constraints associated with local translations.

Let us begin with the case of \tgr\ in $\bR_2$, where the lambda sector of the phase space is gauge fixed. The canonical generator for local translations is given by the standard PG form, see Eqs. (5.1c,d) of Ref.~\cite{mb.mv-2000}:
\begin{align}
\bG_T(\xi)=&-\dot\xi^\m\left(\vth^k{}_\m\pi_k{}^0
  +\frac{1}{2}\om^{ij}{}_\m\pi_{ij}{}^0\right)-\xi^\m\bP_\m\,,       \nn\\
\bP_0:=&\bH_T-\pd_\a D^\a\,,                                         \nn\\
\bP_\a:=&\bcH_\a-\frac{1}{2}\om^{ij}{}_\a\cH_{ij}
             +\pi_k{}^0\pd_\a\vth^k{}_0 +\frac{1}{2}\pi_{ij}{}^0\pd_\a\om^{ij}{}_0\,.
\end{align}
The terms that differ from the corresponding $\fT$ expressions are marked by a bar.

In contrast to the Lorentz generator \eq{5.1}, the translational generator is affected by the structure of the Lagrangian.
Namely, since the \tgr\ Hamiltonians $(\bcH_\orth,\bcH_\a)$ do not depend on the $\fT$ variables $(\phi,\pi_\phi)$, the generator $\bG_T$ cannot be a proper generator in $\fT$ gravity. According to Ref.~\cite{mb.jn-2020}, section 4, the problem can be simply resolved by replacing the \tgr\ Hamiltonians $(\bcH_\orth,\bcH_\a)$ with the corresponding $\fT$ expressions $(\cH_\orth,\cH_a)$, given by
\bsubeq
\be
\cH_\orth=\bcH_\orth+\hu_\phi\pi_\phi\,,\qquad
\cH_\a=\bcH_\a+\pi_\phi\pd_\a\phi\,.
\ee
As a consequence,
\be
P_0=\bP_0+\pi_\phi\pd_0\phi\,,\qquad P_\a=\bP_\a+\pi_\phi\pd_\a\phi\,.
\ee
\esubeq
Then, applying the inverse Castellani algorithm to the $\fT$ gauge generator $G_T$, one can conclude:
\bitem
\item[(c1)] The two sets of constraints, $(\pi_k{}^0,\pi_{ij}{}^0)$ and $(P_0,P_\a)$, are both first class. Since $\cH_{ij}$ is shown to be first class in section \ref{sec5}, it follows that $(\cH_\orth,\cH_\a)$ must also be first class.
\eitem

The above analysis can be straightforwardly extended ``backwards" to the phase space $R_1$. Namely, the gauge generator $G_T$ in $R_1$ is modified by the presence of several additional terms, proportional to the lambda momenta $\pi_{ij}{}^{\a\b}$ \cite{mb.mv-2000}. However, as shown in section \ref{sec4}, these momenta are first class constraints, so that the above conclusion remains valid also in $R_1$.

\section{On the coframe-Lorentz formulation of TG}\label{appC}
\setcounter{equation}{0}

Inspired by some ideas appearing in the recent literature, we discuss certain aspects of the \emph{coframe-Lorentz} form of TG, in which the
Lorentz matrix $\L^i{}_n$ appearing in the pure gauge spin connection $\om^{ij}(\L)\equiv \L^i{}_n\rd\L^{jn}$, is treated as an \emph{independent} dynamical variable.

\prg{1.} In the coframe-Lorentz formalism, with $\vth^i$ and $\L^i{}_k$ as the basic dynamical variables, local Lorentz (LL) transformations, parameterized by $\tL^i{}_k$, are naturally defined by
\be
\hat\vth^i:=\tL^i{}_n\vth^n\,,\qquad \hat\L^i{}_n:=\tL^i{}_m\L^m{}_n\,. \lab{C.1}
\ee
As a consequence, the Lorentz transforms of $\om^{ij}(\L)$ and
$T^i\equiv \rd\vth^i+\om^i{}_n(\L)\vth^n$ are given by
\bsubeq\lab{C.2}
\bea
&&\hat\om^{ij}:=\om^{ij}(\tL\L)
            \equiv\tL^i{}_m\big[\om^{mn}(\L)\tL^j{}_n+\rd\tL^{jm}\big]\,,\\
&&\hat T^i:=\rd\hat\vth^i+\hat\om^i{}_m\hat\vth^m \equiv\hat\L^i{}_m T^m\,.
\eea
\esubeq
Hence, the coframe-Lorentz transformations \eq{C.1} correctly reproduce the well-known PG rules.

In the particular case when $\tL=\L^{-1}\equiv\L^T$, relations \eq{C.2} imply
\bsubeq
\bea
&&\hat\om^{ij}=0\,,           \\
&&\hat T^i=\rd\hat\vth^i\,,
\eea
\esubeq
which describes the pure coframe TG. The Lagrangian (which is a Lorentz scalar valued function of its arguments) can be transformed
to an equivalent form as follows:
\be
L(\vth^i,T^i)\equiv L(\tL^i{}_n\vth^n,\tL^i{}_n T^n)
              \equiv L(\hat\vth^i,\hat T^i)\equiv L(\hat\vth^i,\rd\hat\vth^i)\,.
\ee
Thus, the coframe-Lorentz TG (left) and its pure coframe form (right) are interrelated by an LL tranformation, hence they are physically indistinguishable.

\prg{2.} In the following parts of this appendix, to aid clarity, we will use tilde indicies to refer to the teleparallel frame,\footnote{Unique up to constant Lorentz transformations.} $\vth^\tn=\L_j{}^\tn \vth^j$, which satisfies the Weitzenb\"ock gauge condition (vanishing connection):
\be
\om^\tk{}_\tn=\L_i{}^\tk\left(\om^i{}_j\L^j{}_\tn+\rd \L^i{}_\tn\right)=0\,.
\ee

A deeper understanding of the LL invariance can be obtained in the Hamiltonian approach. Let us begin by defining the canonical momenta associated to $(\vth^i{}_\m,\L^{i\tn})$ by $\pi_i{}^\m:={\pd L}/{\pd T^i{}_{0\m}}$ and
$P_{i\tn}:={\pd L}/{\pd\pd_0\L^{i\tn}}$, which, in view of $T^i{}_{0\a}=\pd_0\vth^i{}_\a+\L^i{}_\tn\pd_0\L^{j\tn}\vth_{j\a}-\nabla_\a\vth^i{}_0$, lead to the 6 primary constraints
\be
\phi_{ij}:= P_{[i}{}^\tn\L_{j]\tn}
                         +\pi_{[i}{}^\a\vth_{j]\a}\approx 0\,.    \lab{C.4}
\ee
Next, we introduce the canonical Hamiltonian
\be
H_c:=\pi_i{}^\a\pd_0\vth^i{}_\a+P_{i\tn}\pd_0\L^{i\tn}-L
    =\pi_i{}^\a (T^i{}_{0\a}+\nabla_\a\vth^i{}_0)-L \,,           \lab{C.5}
\ee
where the last equality follows from \eq{C.4}. Using the definition of $\pi_i{}^\a$, one can eliminate the ``velocities" $T^i{}_{0\a}$ and obtain $H_c$ as a function on the phase space.

The form of the constraints \eq{C.4} indicates that they are first class.
To test this possibility, one could calculate the PB algebra of constraints, but the inverse Castellani algorithm is more simple. It is based on showing that the object $G:=\int \ve^{ij}\phi_{ij} d^3 x$ generates LL transformations of phase space variables $Z$ by $\d_0 Z:=\{Z,G\}$.\footnote{In the next paragraph, we show that calculations based on the naive PBs produce the correct result.} The result takes the expected form, $\d_0\vth^i{}_\a=\ve^i{}_j\vth^j{}_\a$,
$\d_0\L^{i\tn}=\ve^i{}_j\L^{j\tn}$, and similarly for the momentum variables. Hence, $G$ is the correct gauge generator and consequently, $\phi_{ij}$ are indeed first class.

\prg{3.} The Lorentz matrix $\L^i{}_\tn$ has 16 components, but not all of them are independent---they satisfy 10 constraints $\L^\text{T}g\L=g$, so they effectively have only 6 degrees of freedom. To examine the dynamical consequences of this consider the identity
\be
\pd_\m\L^{j\tm}\equiv\mathbb{P}^{j\tm}{}_{l\tn}\,\pd_\m\L^{l\tn}\,,\qquad
\mathbb{P}^{j\tm}{}_{l\tn}:=\frac{1}{2}(\d^\tm_\tn\d^j_l-\L_l{}^\tm\L^j{}_\tn)\,.
\ee
Iterating, one gets $\pd\L=\mathbb{P}\pd\L=\mathbb{P}^2\pd\L$. This suggests that $\mathbb{P}$ is a \emph{projector}, $\mathbb{P}=\mathbb{P}^2$, a property that is easily verified. Then, since $\mathbb{P}$ is self-adjoint,
$\L\pd\L=(\L\mathbb{P})(\mathbb{P}\pd\L)\equiv (\mathbb{P}\L)(\mathbb{P}\pd\L)$.
Thus each $\L$ in the combination $\L\pd\L$ that appears in $\om^{ij}(\L)$ has effectively only six independent (Lorentz) components, the same as the number of independent components of $\om^{ij}(\L)$.
Similarly, the term $P\pd_0\L$ in $H_c$ can be written as $P\pd_0\L=(\mathbb{P}P)(\mathbb{P}\pd_0\L)$, showing that effectively, both $\L$ and $P$ have six independent components each.

The same projection technique could be used in defining the Poisson bracket---but this is not essential---in practice, one has the option of using the more simple, direct 16 component representation. With $N=16+6$ independent Lagrangian variables $(\vth^i,\L^i{}_\tm)$, and generically, with just the diffeomorphism and Lorentz first class constraints $(N_1 = 2\cdot 4 + 6)$ and no second class constraints $(N_2 = 0)$, the number of d.o.f is $N^\star = N-N_1-N_2/2 = 8$, as expected.

\prg{4.} The above considerations allow us to make a few simple
comments on the  literature.\par

Let us start with the work of Blixt et al.~\cite{blixt-2018,blixt-2019}, whose aim is to show that the number of d.o.f in the coframe-Lorentz TG is not affected by fixing the LL symmetry. For that purpose, they introduce a kind of ``instantaneous angular velocity'' variable $a_{ij}=\L_i{}^\tn\pd_0\L_{j\tn}$ and its associated ``canonical momenta" $\hpi^{ij}:=\pd L/\pd a_{ij}$, which satisfy the constraints\footnote{The constraints \eq{C.4} and \eq{C.7} imply $\hpi^{ij}a_{ij}=P_{j\tn}\pd_0\L^{j\tn}$, and consequently $\pi_i{}^\a\pd_0\vth^i{}_\a+\hpi^{ij}a_{ij}=\pi_i{}^\a(T^i{}_{0\a}+\nabla_\a\vth^i{}_0)$, which gives the same form of the canonical Hamiltonian as in \eq{C.5}.}
\be
\Phi^{ij}:=\hpi^{ij}-\pi^{[i\a}\vth^{j]}{}_\a\approx 0\,.           \lab{C.7}
\ee
Although the momenta $\hpi^{ij}$ are not independent of $\pi_i{}^\a$, one cannot conclude that they are pure gauge variables, as the authors claim. Namely, to verify the gauge nature of $\hpi^{ij}$, one needs to know the form of their LL transformations and the related first class constraints. Although the constraints $\Phi^{ij}$ look like they might be first class, the authors
did not address the issue of precisely verifying
that possibility. In particular, they did not explain which variable is the canonically conjugate partner of $\hpi^{ij}$.
In the absence of more convincing arguments, one cannot accept the authors' analysis as a valid Hamiltonian proof of the pure gauge nature of $\L^i{}_k$.

The Hamiltonian analysis of the coframe-Lorentz \tgr\ by Golovnev et al. \cite{golovnev-2021} is to a large extent complete. After defining the $\L$ sector of the phase space by the canonically conjugate pair $(\L^i{}_\tm,P_i{}^\tm)$, they identify three sets of the primary constraints: the standard diffeomorphism constraints $\p_i{}^0$,  the standard Lorentz constraints \eq{C.4}, and  the extra Lorentz constraints $C_{ij}$, stemming from the special structure of \tgr. By recognizing the role of projectors for a proper definition of PBs, they show that the PB algebra of the primary constraints is closed. Their analysis could be easily completed by using the diffeomorphism invariance to conclude that the four Hamiltonian components  are (secondary) first class constraints, see Appendix \ref{appB}. As a consequence, all the constraints of \tgr\ are first class, as expected \cite{mb.in-2000}. However, in the case of $\fT$ gravity, the extra LL symmetry associated to $C_{ij}$ is broken \cite{mb.jn-2020}.


\end{document}